\documentclass[sigconf]{acmart}

\usepackage{balance}
\usepackage{booktabs}
\usepackage{balance}
\usepackage{array}
\usepackage{listings}
\usepackage{enumitem}
\usepackage{graphicx}
\usepackage{subcaption}
\usepackage{multirow}
\usepackage{url}
\usepackage{color, colortbl}
\definecolor{Gray}{gray}{0.9}

\usepackage{amsmath}
\usepackage{mathtools}
\usepackage{commath}
\usepackage{csquotes}

\usepackage[english=usenglishmax]{hyphsubst}

\balance
\newtheorem{definition}{Definition}

\begin{document}

\author{Besnik Fetahu$^1$, Avishek Anand$^1$, Maria Koutraki$^{1,2}$} 
\affiliation{ 
  \institution{$^1$L3S Research Center,  Leibniz University of Hannover, Hannover, Germany}
  \institution{$^2$FIZ-Karlsruhe Leibniz Institute, Karlsruhe Institute of Technology, Karlsruhe, Germany}
}
\email{{fetahu,anand,koutraki}@L3S.de}

\fancyhead{}
\balance
\sloppy

\copyrightyear{2019} 
\acmYear{2019} 
\setcopyright{acmcopyright}
\acmConference[WWW '19]{Proceedings of the 2019 World Wide Web Conference on World Wide Web, WWW 2019, San Francisco, USA.}
\acmBooktitle{Proceedings of the 2019 World Wide Web Conference on World Wide Web, WWW 2019, San Francisco, USA.}
\acmPrice{15.00}

\lstset{basicstyle=\footnotesize\ttfamily}

\title{TableNet: An Approach for Determining Fine-grained Relations for Wikipedia Tables}
\begin{abstract}

Wikipedia tables represent an important resource, where information is organized w.r.t table schemas consisting of columns. In turn each column, may contain \emph{instance values} that point to other Wikipedia articles or \emph{primitive values} (e.g. numbers, strings etc.). 

In this work, we focus on the problem of interlinking Wikipedia tables for two types of table relations: \emph{equivalent} and \emph{subPartOf}. Through such relations, we can further harness semantically related information by accessing related tables or facts therein.  Determining the relation type of a table pair is not trivial, as it is dependent on the schemas, the values therein, and the semantic overlap of the cell values in the corresponding tables.

We propose \emph{TableNet}, an approach that constructs a knowledge graph of interlinked tables with \texttt{subPartOf} and \texttt{equivalent} relations. \emph{TableNet} consists of two main steps: (i) for any  \emph{source} table we provide an \emph{efficient} algorithm to find all \emph{candidate} related tables with \emph{high coverage}, and (ii) a neural based approach, which takes into account the table schemas, and the corresponding table data, we determine with \emph{high accuracy} the table relation for a table pair.

We perform an extensive experimental evaluation on the entire Wikipedia with more than 3.2 million tables. We show that with more than 88\% we retain relevant candidate tables pairs for alignment. Consequentially, with an accuracy of 90\% we are able to align tables with \texttt{subPartOf} or \texttt{equivalent} relations. Comparisons with existing competitors show that TableNet has superior performance in terms of coverage and alignment accuracy.

\end{abstract}

\maketitle

\section{Introduction}\label{sec:intro}

Wikipedia has emerged as one of the most reputable sources on the internet for a wide range of tasks, from question answering~\cite{DBLP:conf/acl/ChenFWB17} or relation extraction~\cite{DBLP:conf/acl/MintzBSJ09}. One of the most notable uses of Wikipedia is on knowledge base construction. 

Well known knowledge bases like DBpedia~\cite{DBLP:conf/semweb/AuerBKLCI07} or YAGO~\cite{DBLP:conf/www/SuchanekKW07} are almost exclusively built with information coming from Wikipedia's infoboxes. Infoboxes have several advantages as they adhere to pre-defined templates and contain factual information (e.g. \emph{bornIn} facts). However, they are sparse and the information they cover is very narrow. For most of the application use cases of Wikipedia, availability of factual information is a fundamental requirement.  

Wikipedia tables on the other hand are in abundance. The current snapshot of Wikipedia contains more than 3.23M tables from more than 520k Wikipedia articles. Tables are rich with factual information for a wide range of topics. Thus far, their use has been limited, despite them covering a broad domain of factual information that can be used to answer complex queries. For instance, for complex queries like \emph{``Award winning movies of horror Genre?''} the answer can be found from facts contained in \emph{multiple tables} in Wikipedia. However, question answering systems~\cite{DBLP:conf/www/AbujabalRYW18} built upon knowledge base facts, in most cases they will not be able to provide an answer or will provide an \emph{incomplete} answer.

The sparsity or lack of factual information from infoboxes can easily be remedied by additionally considering facts that come from Wikipedia tables. A rough estimate reveals that we can generate more than hundreds of millions of additional facts that can be converted into knowledge base triples~\cite{DBLP:conf/wsdm/MunozHM14}. This amount in reality is much higher, if we allow for tables to be aligned. That is, currently, tables are seen in isolation, and \emph{semantically} related tables are not interlinked (i.e. \texttt{equivalent} table relations). Table alignments would allow to access tables fulfilling a specific criteria (e.g. \emph{``List of All Movies''} by different \texttt{Producers}). Additionally, relations that can semantically describe tables as \emph{supersets or subsets} (i.e. \texttt{subPartOf} relations) in terms of classic database \emph{projection} or \emph{selection} functions are missing (cf. Figure~\ref{fig:table_example}), which would enable queries to access semantically dependent tables (i.e.,  \emph{``List of Award-Winning Movies''} and \emph{``List of All Movies''} from a \texttt{Producer}). The presence of such fine-grained relations opens up for opportunities that can be used in question answering, knowledge base construction, and inferences of other facts from the facts that reside in \texttt{equivalent} or \texttt{subPartOf} aligned tables.

Determining the fine-grained table relations is not a trivial task. Table relations are dependent on the semantics of the columns (e.g. a column containing instance values of type \texttt{Country}), the context in which the column appears (e.g. \emph{``Name''} can be an ambiguous column and it can only be disambiguated through other columns in a table schema), cell values etc. Furthermore, not all columns are important for determining the relation between two tables~\cite{DBLP:conf/sigmod/SarmaFGHLWXY12}. Finally, to be able to establish relations amongst all relevant table pairs, requires for efficient approaches that avoid exhaustive computations between all table pairs that can be cumbersome given the extent of tables in Wikipedia.

In this aspect, related work has focused mainly on table retrieval scenarios. The Google Fusion project~\cite{DBLP:conf/sigmod/SarmaFGHLWXY12,DBLP:journals/pvldb/CafarellaHWWZ08} retrieves top--$k$ tables, where the query is a table and the notion of relatedness is in terms of table schemata (specifically subject columns). Recent work~\cite{DBLP:conf/www/ZhangB18} focuses on ad-hoc table retrieval from keyword search. There are two main issues that are not addressed by related work: (i) top--$k$ retrieval does not provide guarantees in terms of coverage, and (ii) the notion of relevance is in terms of \emph{keyword queries}, and there is no distinction between the different \emph{relation types}, specifically \texttt{equivalent} and \texttt{subPartOf} relations.

We propose \emph{TableNet}, an approach with the goal of aligning tables with \texttt{equivalent} and \texttt{subPartOf} fine-grained relations. Our goal is to ensure that for any table, with high coverage we can find candidate tables for alignment, and with high accuracy determine the relation type for a table pair. We distinguish between two main steps: (i) efficient and high coverage table candidate generation for alignment, and (ii) relation type prediction by leveraging table schemas and values therein.

We perform an extensive evaluation of \emph{TableNet} on the entire English Wikipedia with more than 3.2 million tables. Through our proposed approach we are able to retain table pairs that have a relation with a high coverage of 88\%, and correspondingly predict the type of the relation with an accuracy of 90\%.

We make the following contributions in constructing TableNet:
\begin{itemize}[leftmargin=*]
  \setlength{\parskip}{0pt}
  \setlength\itemsep{0em}
    \item we formally define the problem of fine-grained table alignment;
    \item we model tables with fine-grained information relying on column descriptions, instance values, and types and additionally take into account the context in which the columns appear for table alignment;
	\item a large ground-truth for table alignment with coverage guarantees with more than 17k table pairs;
	\item a large scale knowledge graph of aligned tables with more than 3.2 million tables.
\end{itemize}

\section{Related Work}\label{sec:relatedwork}

In this section, we review related work, which we differentiate between three main categories that we describe in the following.

\textbf{Wikipedia Tables.} Bhagavatula et al.~\cite{DBLP:conf/kdd/BhagavatulaND13} propose an approach for finding relevant columns from target tables for a given source table, based on which the table pair can be joined.  Similarly, in our table candidate generation process we employ graph relatedness measures in order to generate relevant table pairs for alignments. Nonetheless, our objectives are on finding tables for alignment that are semantically similar. As such our criteria for alignment does not adhere to their \textit{join} definition where the objective is to construct a new table as a result of the joined tables. Additionally, for our task of fine-grained alignment, we argue that relying only in specific columns is not sufficient for alignment. The semantics of a table cannot always be defined from a single column, but rather the context in which the column occurs.

\textbf{Web Tables.} The work carried in the project \textit{Google Fusion Tables}~\cite{DBLP:journals/pvldb/CafarellaHWWZ08,DBLP:conf/sigmod/SarmaFGHLWXY12,DBLP:journals/pvldb/VenetisHMPSWMW11,DBLP:conf/sigmod/GonzalezHJLMSSG10} represents one of the most significant efforts in providing additional semantics over tables, and to the best of our knowledge, only some of the works carried in this project are most related to our work, against which we provide an optimal comparison~\cite{DBLP:conf/sigmod/SarmaFGHLWXY12}.

Carafella et al.~\cite{DBLP:journals/pvldb/CafarellaHWWZ08} propose an approach for table extraction from Web pages and additionally provide a ranking mechanism for table retrieval. An additional aspect they consider is the schema auto-completion for some input column, where they recommend other columns that would fit contextually to generate a ``complete'' schema. Our aim is different here, while we aim at providing more fine-grained representations of columns in a table schema, our goal is to use such information for the task of table alignment.

Das Sarma et al.~\cite{DBLP:conf/sigmod/SarmaFGHLWXY12} propose an approach for finding related tables, where as relatedness they consider two cases: (i) \emph{entity complement} and (ii) \emph{schema complement}. For (i), the task is to align tables that have the same table schemas, however, with complementary instances. This case can be seen as applying a \emph{selection} over some table that has the union of instances from both tables. In (ii), the columns of a target table can be used to complement the schema of a source table, with the precondition that the instances (from \emph{subject} columns) are the same in both tables. This case is seen as a \emph{projections} operation over some table with the same \emph{selection} criteria, thus, resulting in the same set of instances.

Our work is related to the case of \emph{entity complement}, where the authors compute the \emph{schema similarity} between two tables in order to decide if a table can be considered for complementing the instances in another table. The similarity of the schemas is considered as a max-weight bipartite matching approach, with weighted established between the column in the disparate schemas, and edge weight being the \emph{string similarity} between columns and jaccard similarity between the column types (established from the values in a column through the \texttt{WebIsA} database). 

Despite the fact that this approach is unsupervised, we adapt it such that we find the best threshold of the max-weight matching score between two schemas, and consider tables to be either \emph{aligned} or \emph{not-aligned}. We show that our approach outperforms the most closely related work from Google Fusion.

The aforementioned works rely on a structured table representation based on work by Venetis et al.~\cite{DBLP:journals/pvldb/VenetisHMPSWMW11}. The columns in a table are labelled based on a \texttt{isA} database, which consists of instances and the associated labels mined from a large Web corpus (e.g. \emph{capital cities} for a column containing \emph{Paris, Berlin}, etc.). In our case, we deal with Wikipedia tables in which instances are linked to Wikipedia articles, thus, we opt for using the Wikipedia category taxonomy and the additional information coming from knowledge bases for inferring a structured representation of a table schema, respectively for the individual columns in a table. Wikipedia categories are much richer than the \texttt{isA} database used in \cite{DBLP:journals/pvldb/VenetisHMPSWMW11}, which is flat, contrary to the categories which represent a taxonomy, thus, allowing us to leverage from coarse to fine grained information about columns.

\textbf{Table Annotation.} Work on table annotation \cite{DBLP:journals/pvldb/LimayeSC10,DBLP:conf/semweb/BhagavatulaND15} focus specifically on linking cell values with entities, and columns with entity types that best describe the values in a column. We see these works as complementary, in which case we can employ them to further enrich tables where the cell values are not already linked to existing Wikipedia entities. 

Another line of work is proposed by Mun\~oz et al.\cite{DBLP:conf/wsdm/MunozHM14}. In this work, Wikipedia tables, specifically table rows are used as an input for generating RDF triples, where the relations between two columns correspond to properties extracted from a target knowledge base like DBpedia. Slightly related to \cite{DBLP:conf/wsdm/MunozHM14} is the work by Ritze et al.~\cite{DBLP:conf/wims/RitzeLB15}, where the authors propose a system called \emph{T2K Match}. T2K matches Web tables into a target knowledge base, that is, the columns are described in terms of classes from the target KB. The works in \cite{DBLP:conf/wims/RitzeLB15,DBLP:conf/wsdm/MunozHM14} can be seen as complementary and they could be used as additional information for the table alignment task.

\textbf{Schema Matching.} In the database community, there has been extensive research in schema matching~\cite{rahm2001survey,DBLP:conf/dexa/NunesCCFLD13,DBLP:conf/edbt/KoutrakiPV16,DBLP:conf/esws/KoutrakiPV17}. However, works in schema matching tackle the problem of mapping individual columns from two database table schemas, whereas, in our case, the column alignments from two table schemas are only intermediary input into determining the actual fine-grained relation between two tables.

\section{Preliminaries and Overview}\label{sec:problem}

\begin{figure*}[ht!]
	\centering
	\includegraphics[width=.80\textwidth]{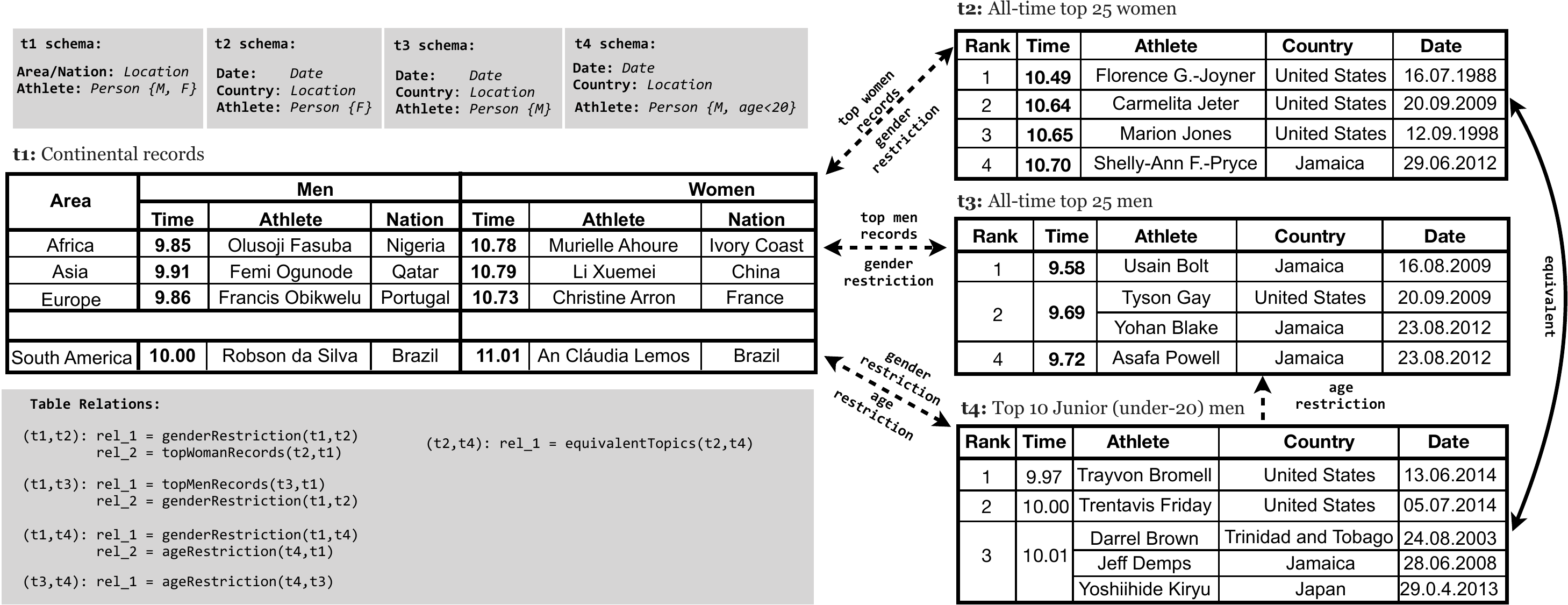}
	\caption{Table alignment example with \texttt{subPartOf} (dashed line), and \texttt{equivalent} relations (full line). \texttt{subPartOf} can be explained in terms of \emph{age restriction} or \emph{gender restriction}, whereas \texttt{equivalent} relation represents topically similar information.}
	\label{fig:table_example}
\end{figure*}

\subsection{Terminology} 
We consider Wikipedia articles $A=\{a_1,\ldots,a_n\}$; each article $a$ is associated with a set of Wikipedia categories $\Psi_a=\{\psi_1,\ldots,\psi_n\}$. From all the categories we induce the category graph $\Psi$, which consists of  \emph{parent} and \emph{child} relations between categories $\psi_i \text{\texttt{ childOf }} \psi^p$. The parent/child relations allow us to establish a hierarchical graph in $\Psi$. The level of a category is denoted by $\lambda_{\psi}$. 

Next, we define the tables from an article $a$ as $T_a=\{t_1,\ldots,t_n\}$. A table $t$ consists of a \emph{table schema} (or \emph{column header}) that we define as $C(t)=\{c_1,\ldots c_n\}$. Each column consists of a \emph{textual description} and the set of all values $c_i=\langle desc, \{v_i^1,\ldots,v_i^n\}\rangle$ assigned to the corresponding column cells in the table rows $t_i(r)=\{r_i^1,\ldots,r_i^n\}$. More specifically, the cell value that is attributed to a specific \emph{row} and \emph{column} is indicated by $v_i^k$, where $k$ is the row $r^k$ and $i$ is the column $c_i$. Cell values can point to existing articles in Wikipedia, that is $v_i^k= \langle a_k\rangle$, which we will refer to as \emph{instance values}, or \emph{primitive values} in cases of text, numbers, dates etc.

From the extracted tables $T=\{t_1,\ldots,t_n\}$ from $A$, we define two fine-grained types of relations between a table pair $\langle t_i, t_j\rangle$: (i) $t_i \vDash t_j$ where $t_j$ is considered to be \emph{semantically} a \texttt{subPartOf} of $t_i$, and (ii)  $t_i \equiv t_j$ where $t_i$ and $t_j$ are \emph{semantically} \texttt{equivalent}. We indicate the presence of relation with $r(t_i, t_j)\neq \emptyset$, and in the next section we precisely define the table relations.

\subsection{Table Alignment Task Definition}\label{subsec:prob_definition}

In this section, we define the task of table alignment, and provide the definition for the fine-grained table relations.

\paragraph{\textbf{Table Alignment}} From the generated table pairs in the previous step, the task is to determine the \emph{relation type} between any table pair $r(t_i,t_j)$ from the article pair $\langle a_i, a_j\rangle$. The relation types can be either \texttt{subPartOf}, \texttt{equivalent} or \texttt{none} (in case $r(t_i, t_j)=\emptyset$).

\begin{definition}[subPartOf] For a table pair $r(t_i, t_j)=\{\text{\texttt{subPartOf}}\}$ holds if the schema $C(t_i)$ can subsume either at the \textbf{data value} (i.e. \textbf{cell-value}) or \textbf{semantically} the columns from $C(t_j)$ (cf. Figure~\ref{fig:table_example}), that is, $C(t_i)\supseteq C(t_j)$. 
\end{definition}

\begin{definition}[equivalent] For a pair  $r(t_i, t_j)=\{\text{\texttt{equivalent}}\}$ holds if both table schemas have \textbf{semantically similar} column representation (cf. Figure~\ref{fig:table_example}), that is, $C(t_i)\approx C(t_j)$.
\end{definition}

With the notion of \emph{semantic similarity} we refer to cell \emph{instance} values, whose similarity can be assessed at the Wikipedia category level, e.g. the column $c=\langle\text{\emph{``Country''}, \{\text{\emph{Germany}}, \text{\emph{USA}}, \ldots \}}\rangle$ contains values of type \texttt{Location}, thus $c$ will be semantically similar to any column whose values are topically similar, despite the fact that at data level those values will not overlap. Similarly, is the case for the column descriptions, where \emph{``Nation''} and \emph{``Country''} usually refer to the same type of columns.

\subsection{TableNet Overview}\label{sec:approach}

TableNet operates in a manner such that for any given \emph{source} Wikipedia article $a_i\in A$, first, we generate article candidate pairs $\langle a_i, a_j\rangle$ ($a_j \in A$ and $a_i\neq a_j$) where the tables from the pair are likely to have an alignment relation (i.e. $r(t_i, t_j)\neq \emptyset$, where $t_i \in T(a_i)$ and $t_j \in T(a_j)$), which corresponds to the second step in TableNet. In the following, we describe the two main steps of TableNet:
\begin{enumerate}
	\setlength{\parskip}{0.2em}
	\setlength\itemsep{0.2em}
	\item Article candidate pair generation
	\item Table alignment
\end{enumerate}

\section{Article Candidate Pair Generation}\label{subsec:table_extraction}

In the article candidate generation step, we address the problem of determining article pairs, whose tables will result in a table relation. In this case, we require our article candidate generation process to fulfill two main properties. First, we need to  minimize the amount of \emph{irrelevant article pairs}, whose tables do not result in a table relation. Second, the filtering out of article pairs (from the first property) should not affect the coverage in terms of retaining \emph{relevant article pairs}, whose tables result in a table relation. 

Thus, we define a function as shown in Equation~\ref{eq:td_table_align_pairs} which provides optimal coverage of relevant article pairs, and at the same time minimizes the amount of irrelevant pairs.
\begin{equation}\label{eq:td_table_align_pairs}
A\times A \rightarrow r(t_i, t_j), \text{ where } t_i \in T(a_i) \wedge t_j \in T(a_j), \text{ and } a_i \neq a_j 
\end{equation}

This step is necessary as \emph{naive} approaches which would enumerate over all article pairs to ensure maximal coverage result in a combinatorial explosion ($n!$ where n is the number of articles). Thus, in our candidate generation approach we circumvent this issue by defining features that fulfill the following desiderata:
\begin{itemize}[leftmargin=*]
    \item For an article pair whose tables will result in a table alignment, we expect the articles to be \emph{semantically} or \emph{topically} similar, which can be captured through the article abstracts or their category associations.
    \item For any two tables, whose parent articles fulfill the first criterion, we expect to find notions of similarity in terms of their schemas, such as column names
\end{itemize}

With these desiderata in mind, we define features that either operate at the article pair level or table level, and use them in two ways: (i) filter out irrelevant article pairs, and (ii) employ the features in a supervised manner to further filter out such pairs.

\begin{table}[ht!]
	\centering
	\scalebox{0.75}{
	\begin{tabular}{l l p{5.5cm} p{1.5cm}}
		\toprule
		& \emph{feature} & \emph{description} & \emph{group}\\
		\midrule
		$f_1$ & \emph{tfidf} & \emph{tfidf} similarity between abstracts & \multirow{3}{*}{\emph{abstract}} \\
		$f_2$ & \emph{d2v} & \emph{doc2vec} similarity between abstracts & \\
		$f_3$ & \emph{w2v} & \emph{avg.} word2vec abstract vectors similarity & \\
		\midrule
		$f_4$ & $sim(\Psi_{a_i}, \Psi_{a_j})$ & similarity in embedding space between $\Psi_a$ and $\Psi^p_a$ categories for the article pair & \multirow{5}{2cm}{\emph{$\Psi$ \& KB}}\\
		$f_5$ & $\bigcap\limits_{a \in \langle a_i, a_j\rangle}\Psi_a$ & direct and parent categories overlap & \\
		$f_6$ & $sim(a_i, a_j)$ & embedding similarity of the article pair& \\
		$f_7$ & \emph{type} & type overlap & \\
		
		\midrule
		$f_8$ & $sim(\psi_i, \psi_j)$ & column title similarity ($f_8^l$) and column distance ($f_8^d$) between the schemas in a table pair & \emph{tables}\\
		$f_9$ & $\norm{\gamma(\psi_i)-\gamma(\psi_j)}$ & category representation similarity $\gamma$\\
		\bottomrule
	\end{tabular}}
	\caption{Article candidate pair features.}
	\label{tbl:candidate_feature_list}
\end{table}

\subsection{Features} Table~\ref{tbl:candidate_feature_list} provides an overview of all the similarity features for the article candidate pair generation step.

\textbf{Article Abstract.} Wikipedia article abstracts contain a summary of the article, containing the most important information. For a \emph{relevant article pair}, we expect that the abstracts will overlap in terms of abstract, specifically on \emph{keywords}. For example, \texttt{Ariane Friedrich} and \texttt{Teddy Tamgho}, whose tables are in \texttt{equivalent} relation, both  contain snippets indicating that the corresponding persons are \emph{\textbf{athletes}} and \emph{\textbf{jumpers}}. This \emph{topical similarity} is in line with the definitions of table relations in Section~\ref{sec:problem}.

The features in Table~\ref{tbl:candidate_feature_list} in the \emph{abstract} group capture exactly such \emph{topical} similarities. Feature $f_2$ computes a \emph{doc2vec}~\cite{DBLP:conf/icml/LeM14} embedding for each article and measures the cosine similarity between those embeddings. Doc2Vec embeddings have the advantage that they take into account a broader context when compared to standard word2vec~\cite{mikolov2013distributed}. Additionally in $f_3$, we compute an \emph{average embedding} from \emph{word2vec} embeddings from the tokens in the abstract. We use GloVe pre-trained embeddings  Wikipedia~\cite{DBLP:conf/emnlp/PenningtonSM14}. Finally, since none of the embedding can account for the \emph{salience of tokens} in an abstract, we additionally compute the cosine similarity of the \emph{tf-idf} (feature $f_1$) vectors from  the article abstracts.

\textbf{Categories \& Knowledge Bases.} Wikipedia categories are indicators that two articles are semantically similar in either \emph{temporal}, \emph{geographical}, or \emph{topical} dimension (categories are created to indicate either \emph{temporal} or \emph{geographical} grouping, or \emph{topical} similarity). Since articles are associated manually to categories, this association is prone to noise. As we show later in our evaluation, if we consider as pairs only articles assigned to the same categories, the resulting coverage will be low.

To circumvent this problem, we compute \emph{graph embeddings} for Wikipedia categories based on the category graph $\Psi$. We use different embedding approaches like RDF2Vec~\cite{DBLP:conf/semweb/RistoskiP16} and Node2Vec~\cite{DBLP:conf/kdd/GroverL16}. This allows us to elevate the category comparisons from the link based structure in $\Psi$ to the embedding space, and consider as candidates articles whose categories are close in the embedding space. 

As features we consider the cosine similarity between the \emph{directly associated categories} for an article pair, and additionally their \emph{parent categories}. We also consider the similarity of articles in the same embedding space (Equation~\ref{eq:cosine_sim}), and the jaccard similarity in terms of types in DBpedia.
\begin{equation}\label{eq:cosine_sim}\small
	sim(a_i, a_j) = \frac{emb(a_i)\cdot emb(a_j)}{\norm{emb(a_i)} \norm{emb(a_j)}}
\end{equation}

\textbf{Tables.} The article pair features capture a coarse grained similarity for the tables in the articles themselves. We compute a set of light-weight features between the tables schemas for tables $t_i\in a_i$ and $t_j\in a_j$ from an article pair. The similarity corresponds to the \emph{average word embedding} of the column description for two columns in the schemas $C(t_i)$ and $C(t_j)$. We consider only the highest matching column as measured in  Equation~\ref{eq:table_column_features}. In addition to the similarity, we capture also the \emph{positional index} difference between the highest matching columns, and the maximal matching in terms of a \emph{column representation} which we explain below.
\begin{align}\label{eq:table_column_features}\small
		\max_{c_l \in C(t_j)} sim(c_k, c_l), 
\end{align}
where $sim(c_k, c_l)$ is computed similarly as in Equation~\ref{eq:cosine_sim}.

\textbf{Column-Representation. } In case a column $c_i$ consists of instance-values, we compute a  representation for $c_i$ based on the attributes associated with the instances $\{v_i\}$ (where an instance points to a Wikipedia article), e.g. \emph{$v_i^1$=``George Lucas'}  \textbf{\emph{bornIn}} \emph{``Modesto, California, U.S.''}. More specifically, since there may be multiple instances $|v_i|>1$, we find the \emph{lowest common ancestor} $\psi_{L}$ category from $\{v_i\}$ by following the article-category associations in $\Psi$. This provides an abstraction over the values and is seen as a \emph{type} of instances in $\{v_i\}$. By considering $\psi_{L}$ instead of the individual $\{v_i\}$, we can summarize the column representation in terms of the most discriminative attributes in overall for $\psi_L$. In this way, for a table pair, we compare the column representations, and in cases of a high match we assume the columns to be semantically similar. 

The representation of $c_i$ is computed as in Equation~\ref{eq:category_attribute_weight}. We weigh the importance of \emph{attributes} based on how discriminative they are for $\psi_L$, e.g. an attribute associated with articles directly belonging to category $\psi_L$ are exclusive for $\psi_{L}$, and thus are weighted high. For an attribute $p$, the weight for $\psi_{L}$ is computed as following:
\begin{equation}\label{eq:category_attribute_weight}\small
	\gamma(p,\psi_{L}) = \frac{\lambda_{\psi_L}}{\max\lambda_{\psi}} * 
	\left(-\log\frac{|\bigcup{o}|:  \forall \langle a, p, o\rangle \wedge a\in \psi_L}{|o|: \forall \langle a, p, o\rangle \wedge a\in \psi_L}\right)
\end{equation}
where, the first part of the fraction weighs $p$ by taking into account the level of $\lambda_{\psi_L}$ and the deepest category where $p$ is present in a target KB $\max\lambda_{\psi}$. $|\bigcup{o}|$ represents the number of distinct values assigned to attribute $p$ from $a\in \psi_L$, whereas $|o|$ is the total number of assignments of $p$ in $\psi_L$.

Through $\gamma(\psi_L)$ we capture the most \emph{important} and \emph{descriptive} attributes for a column $c_i$. For two columns in two table schemas, a high similarity $\norm{\gamma(\psi_i)-\gamma(\psi_j)}$ is an indicator that the columns are semantically similar, which we use as a \emph{table feature}.

\subsection{Filtering \& Classification} 
We use the features in Table~\ref{tbl:candidate_feature_list} in two ways: (i) filter out article pairs that are \emph{unlikely} to yield a table relation, and (ii) train a supervised model and classify article pairs as either \emph{relevant} or \emph{irrelevant}. 

\paragraph{\textbf{Filtering.}} 

We consider a \emph{conjunction} of filtering criteria based on empirically evaluated thresholds for the individual features. Our main goal is to retain a high coverage of \emph{relevant article pairs}, and at the same time filter out drastically \emph{irrelevant pairs}. For thresholds we consider the \emph{mean} value of a particular feature. This ensures that for a pair, if the score is below the mean value, that is an indicator that the pair is unlikely to yield any table relation. In Section~\ref{sec:evaluation} we show that we are able to drastically reduce the number of pairs by simply applying such thresholds.

\paragraph{\textbf{Classification.}} From the remaining pairs we train a classification model and classify pairs as being either \emph{relevant} or \emph{irrelevant}. We consider as positive instances all table pairs from the article pair $\langle a_i, a_j\rangle$ which have \emph{at least one} table relation, i.e, $r(t_i,t_j)\neq \text{\texttt{none}}$, where $\exists (t_i \in a_i \wedge t_j \in a_j)$.

For classification we use Random Forests (RF)~\cite{breiman2001random}. RFs allow to set \emph{minimal amount of samples} that are allowed for a node in the tree to be split. This has direct implications in the accuracy of a classifier, however, this allows us  to retain high coverage. Setting this number high makes the leafs of the different trees to be impure containing relevant and irrelevant article pairs. Our classifier will classify such impure leafs as relevant, at the cost of accuracy, however, in this way we retain a high recall. Section~\ref{sec:evaluation} shows that we can maintain high coverage of relevant pairs and at the same drastically reduce the amount of irrelevant pairs.

\section{Table Alignment}\label{subsec:table_alignment}
TableNet is a bidirectional recurrent neural network (RNN), which for any table pair $r(t_i, t_j)$ learns to classify them into one of the relations \texttt{equivalent}, \texttt{subPartOf}, or \texttt{none}. For a model to accurately align table, the order of columns in their schemas needs to be taken into account. Additionally, the matching columns in the two schemas need to fulfill two main criteria: (i) \emph{the context} in which the columns occur needs to be semantically similar, and (ii) \emph{the positions} in which the columns appear needs to be comparably similar~\cite{DBLP:conf/sigmod/SarmaFGHLWXY12}. 

\begin{figure}[ht!]
	\centering
	\includegraphics[width=1.0\columnwidth]{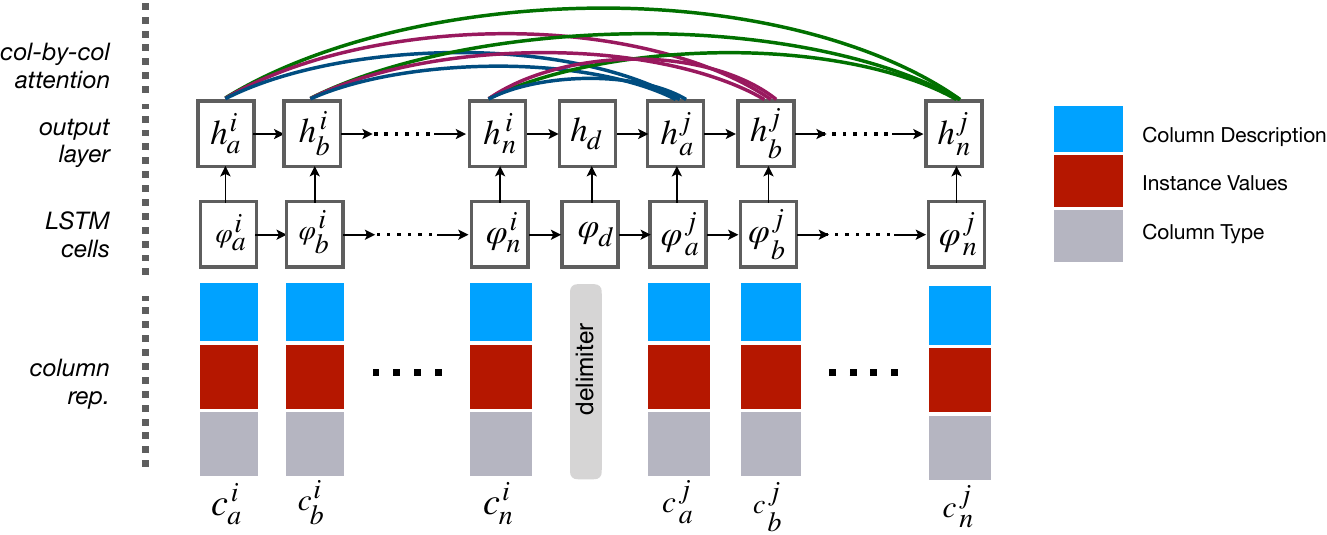}
	\caption{TableNet uses \emph{BiLSTM}s to encode the tables as a sequence of columns. Each column can be represented in terms of its \emph{description}, \emph{instance values}, and its \emph{column-type} (indicated by the different colors). The \emph{column-by-column} captures soft-alignments between columns in the corresponding table schemas. (better viewed in color)}
	\label{fig:alignment_neural_model}
\end{figure}

Figure~\ref{fig:alignment_neural_model} shows an overview of the proposed alignment model. In the following we describe in details the means of representing tables, and the proposed architecture for the alignment task.

\subsection{Table Representation} How we represent columns is key towards an accurate alignment model. A column in table schema consists of the following information $c_i = \langle desc, \{v^1_i, \ldots v^n_1\}\rangle$ (see Section~\ref{sec:problem}).

\textbf{Column Description.} For a column $c$ its description is a strong indicator of the cell-values $v_i$. We represent the column description tokens based on their word embeddings, specifically we use pre-trained Glove word embeddings~\cite{DBLP:conf/emnlp/PenningtonSM14}. In the case of multiple tokens, we average the respective word embeddings.

One disadvantage with this representation is that column descriptions can be ambiguous. For instance, \emph{``Title''} can refer to various different values, e.g. \texttt{Movies, Books} etc. Thus, relying solely on column titles is not optimal for the table alignment process.

\textbf{Instance--Values.} In case $c_i$ contains \emph{instance values}, we represent $c_i$ through the average embeddings of the individual cell-values $v_i$ based on pre-computed graph embeddings. In our experimental evaluation, we used \emph{node2vec} embeddings~\cite{DBLP:conf/kdd/GroverL16}, which we trained on the Wikipedia \emph{anchor graph}\footnote{The anchor graph consists of nodes (Wikipedia articles and categories), while the edges correspond to the anchor text and the category-article associations.}.

The combination of column \emph{description} and \emph{instance values} improves the representation of a column and reduces its ambiguity. This can be mostly attributed to the cases where $c_i$ consists of more than one single instance, whereby by averaging the embeddings we establish the context in which such values can appear.

\textbf{Column--Type.} Representing the columns based solely on instance values poses a risk of biasing the column representation towards articles that are often linked together in the Wikipedia anchor graph, and thus it may ignore the topic information that is present in such articles based on their category associations. 

Hence, for columns that contain instance values, we additionally represent it through its \emph{type} or \emph{category}. That is, for all the instance values in $v_i$ for $c_i$, we extract their lowest common ancestor category from $\Psi$. A similar idea was employed by Das Sarma et al.~\cite{DBLP:conf/sigmod/SarmaFGHLWXY12}, where the columns are represented by the topics they are most often associated with, an information extracted from Google's query logs (e.g. \emph{India} \texttt{Asian country}). Similar as for \emph{instance-value} representation, here too, we represent the LCA category through graph embeddings. As we will see later in our experimental setup (see Section~\ref{sec:setup}), since we we only ensure that $\Psi$ is a consistent hierarchical graph, in cases where the LCA categories can be more than one, then we average their corresponding representations.

\subsection{Table Alignment Architecture} 

In our model, we differentiate the input from the different tables through a \emph{delimiter}. The model is an adoption of the one proposed by Rockt\"aschel~\cite{rocktaschel2016reasoning} in the task of textual entailment. For a table pair $r(t_i, t_j)$ the model predicts the relation $r\in\{$ \texttt{equivalent}, \texttt{subPartOf}, \texttt{none} $\}$.

The alignment model corresponds to an RNN with LSTM cells\cite{hochreiter1997long}, in that we read the sequence of columns for the table pair in both directions. Additionally, on top of the output layer from the RNN model, we compute a \emph{column-by-column} attention, which helps us generate soft-alignments between columns in the  table schemas, and thus further improve the alignment accuracy. In the following we describe the encoding of the column tables, and the intuition behind the attention mechanism.

\paragraph{\textbf{Table Encoding.}} Since we have two separate tables, a precondition for accurate alignment is the encoding of the sequence of columns. Our model provides a \emph{conditional encoding}, in that it first reads the columns from $C(t_i)$, then the cell state $c_d$, which is initialized with the last state of $t_i$ (in this case $c_n^i$) is used to conditionally encode the sequence of columns in $C(t_j)$. 

The advantage of the conditional encoding is that by encoding table $t_j$ with initial cell state that corresponds to the last column cell state of $t_i$, we bias the model to learn encodings that are optimal for the task of table alignment. That is instead of trying to encode all columns, it will learn to encode the columns of $t_j$ such that it can best predict the relation for the table pair. Since we have a bidirectional LSTM, we encode in a similar fashion the table $t_i$ by conditioning it on the last state of $t_j$.

\paragraph{\textbf{Attention Mechanism.}} In our case, for a table pair $r(t_i, t_j)$ to be aligned with either \texttt{equivalent} or \texttt{subPartOf} relation, we expect that the most important columns in each of the tables to have their corresponding matches in the respective schemas. This follows the intuition that not all columns in a table are equally important~\cite{DBLP:conf/sigmod/SarmaFGHLWXY12}.

In this case, if we use the last cell state of the encoded table pair for classification, we enforce the model to weigh equally all the columns for determining the relation type for the table pair. Furthermore, for larger tables, the last cell state is expected to capture the information from all the previous states. A common workaround in such cases is to consider RNNs with attention mechanism~\cite{bahdanau2014neural}. In such cases the models are able to attend over all sequences with a specific \emph{attention weight}. This has the advantage in that classification task is not carried solely based on the last state in a sequence, but instead the sequences are weighed based on their importance.

\textbf{Column-by-Column Attention.} In TableNet, we employ a more sophisticated attention mechanism, which addresses several disadvantages from the global attention mechanism~\cite{bahdanau2014neural}. The \emph{column-by-column} attention mechanism works as following. After having encoded the last column from $t_i$, we process the columns in $t_j$ individually and generate the attention weights w.r.t the columns in $t_i$. As a consequence, for each column in $t_j$ we generate soft alignments to highest matching columns in $t_j$. After having processed all the columns in $t_j$ and computing the corresponding attention weights (the upper part in Figure~\ref{fig:alignment_neural_model}), for classification of the table pair $r(t_i, t_j)$ we will use a non-linear combination of the weighted representation of the last column $c_n^j$ in $t_j$. We use \emph{softmax} classification function for determining the label for $r(t_i, t_j)$.

The advantages of the \emph{column-by-column} attention, is that it allows to encode all the desired table relation semantics, and additionally not enforce for two tables to have the same set of columns, given that not all columns are important for alignment. Thus, the alignment model in TableNet has the following properties:
\begin{itemize}[leftmargin=*]
	\item we can distinguish between columns from the different table schemas $C(t_i)$ and $C(t_j)$, 
	\item for each column in $C(t_j)$ we can compute alignment weights to the column in $C(t_i)$ which function as \emph{soft alignments} between columns in the respective schemas, and 
\end{itemize}

\section{Experimental Setup}\label{sec:setup}

Here we describe the experimental setup for evaluating \emph{TableNet}. First, we introduce the evaluation datasets, and then describe the setup for: (i) candidate generation, and (ii) table alignment.

The evaluation datasets and the code developed for all stages of TableNet are available for download\footnote{\url{https://github.com/bfetahu/wiki_tables}}.

\subsection{Datasets}\label{subsec:datasets} 

The main dataset in our experimental setup is Wikipedia. We use the entire set of Wikipedia articles from the snapshot of \texttt{20.10.2017}, with 5.5 million articles. Additionally, we use the Wikipedia categories, with nearly 1 million categories.

\subsubsection*{Wikipedia Tables.} We extract tables from the HTML content of Wikipedia articles. 
From the entire snapshot of Wikipedia, only 529,170 Wikipedia articles contain tables. This resulted in a total of 3,238,201 Wikipedia tables. 
On average there are \textbf{6 tables per article} with an average of \textbf{6.6 columns}, and with an average of \textbf{10 rows} per table.

In more details, if we consider the composure of columns in the table schemas, more than 20\% of columns in total consist of cell-values that are \emph{instances} (see Section~\ref{sec:problem}). Furthermore, if we consider the number of tables that contain columns with instance values, this number is significantly higher with 85\%. This shows that in the vast majority of cases, we can represent tables, specifically  the columns with highly rich semantic representations.

\subsubsection*{Wikipedia Categories.} The category graph $\Psi$ consists of nearly 1M distinct categories, organized in a \emph{parent-child} graph. However, there are two main issues with using $\Psi$ as is. First, it contains cycles, and second, categories are not depth-consistent, that is, the parents of a category do not belong to the same depth in $\Psi$. 

We resolve these two issues, by first breaking any cycle in $\Psi$, and establish a depth-consistent graph s.t. for every Wikipedia category we remove any edge to its parents ($\psi \text{\texttt{ childOf }} \psi_p$), where the level of the parent category $\lambda_{\psi_p} < \max_{\psi'\in \psi_p}\lambda_{\psi'}$, where with $\Psi_p$ we denote all the parent categories of $\psi$. Removing such edges does not incur any loss in terms of \emph{parent-child} relations between categories, as such categories can be reached through intermediate categories in $\Psi$. This process is performed iteratively from the root category until we have reached the leafs of $\Psi$. 

\subsection{Table Alignment Ground-Truth}\label{subsec:table_alignment_gt}

We are the first to generate a large scale ground-truth for table alignment. Additionally, we are the first to distinguish between \emph{fine-grained relations}, and additionally provide \emph{coverage} guarantees for any given table in terms of its relations.

Our ground-truth consists of a sample of 50 source Wikipedia articles from which we construct article candidate pairs. Since the \emph{naive} approach would generate 26.5M pairs, we  apply a set of \emph{filtering keywords} to filter out irrelevant article pairs. We filter articles by checking if a keyword appears \emph{anywhere} in the article's content. The filtering keywords are chosen to fulfill two main criteria: 
\begin{itemize}[leftmargin=*]
	\item \emph{keywords are \textbf{generic}, in order not to filter out relevant pairs},  
	\item \emph{topical keywords are \textbf{broad} s.t they can capture both coarse/fine grained topics (e.g. ``athletics'' vs. ``jumper'').}
\end{itemize}

We manually inspect a random sample of pairs that are filtered out, and assess if we remove pairs that should be considered relevant, and consequentially refine the keywords. For article pair that remain after filtering, we check if they can be seen as \emph{false positives} and similarly refine our filtering keywords to remove such cases. We iteratively apply the refine and filtering steps, until we are left with an initial set of article pairs that we deploy for evaluation through crowdsourcing. For the remainder of article pairs, we construct all table pairs and rely on crowdsourcing to assess the table relations. Table~\ref{tbl:candidate_pairs} shows the stats for the three filtering iterations w.r.t the 50 source articles. 

From the resulting 3.7k pairs, we have a set of 17k table pairs which we evaluate by means of crowdsourcing.

\begin{table}[h!]
\centering

\begin{tabular}{l l l l}
\toprule
\emph{all pairs} & \emph{iter-1} & \emph{iter-2} & \emph{iter-3}\\
\midrule
26.5M & 416506 ($\blacktriangledown 63\times$) & 10701 ($\blacktriangledown 38\times$) & 3702 $(\blacktriangledown 2.9\times)$\\
\bottomrule
\end{tabular}
\caption{For 50 random source articles we applied three iterations of refine and filtering steps based on manual inspection. The reduction shows the factor with which the filtering reduces the pairs between each consecutive step is w.r.t the previous step. The final set of article pairs, whose table pairs we evaluate through crowdsourcing is 3.7k article pairs.}
\label{tbl:candidate_pairs}
\end{table}

\subsubsection{Evaluation Protocol} 
The table alignment can be of three categories: (i) \texttt{equivalent}, (ii) \texttt{subPartOf}, and (iii) \texttt{none}. To get reliable judgments, we provide detailed instructions and examples to the crowdworkers on how to determine the correct alignment relation.  We guide the crowdworkers through the steps below:
\begin{enumerate}[leftmargin=*]
	\item Find \emph{important} columns that can be interpreted in isolation from the remaining columns in a table schema (such columns are known also as \emph{subject columns}).
	\item Find matching columns in the two tables (that can be considered abstractly as \emph{join keys}), where a match is considered if the columns contain \emph{topically} similar information or the \emph{same} information, and that their column descriptions are matching (e.g. \emph{Nation} is equivalent to \emph{Country}).
	\item If the previous two conditions are met, a table is considered as \texttt{equivalent} if the two tables contain similar or the same information, and where for the important columns in a table there are corresponding columns in the candidate table. 
	\item The alignment \texttt{subPartOf} holds if one of the tables is a super set in the sense that it \emph{contains} the information contained in the other table, or if \emph{semantically} it is the superset of the other table (e.g. \emph{``List of All Movies''} vs. \emph{``List of Award Winning Movies''} for the same movie director). Alternatively, this can be seen as a table generated as a result of a selection over the superset table.
\end{enumerate}

\subsubsection{Evaluation Efforts}

We ensure the quality of the labeling process by using only the highest level workforce in FigureEight. We follow guidelines on how to avoid unreliable workers by establishing a set of \emph{test questions} that we generate manually~\cite{DBLP:conf/chi/GadirajuKDD15}. Every crowdworker needs to pass them successfully with an accuracy of above 70\%. Finally, if a crowdworker takes time less than an estimated  \emph{minimum amount of time} for completing the task, we discard their judgments. 

 On average, it took \textbf{2 mins} for the crowdworkers to judge 5 table pairs. This resulted in a total of \textbf{1,162 hours of work} for the entire ground-truth, for which we payed crowdworkers according to the minimum wage in Germany.

\subsubsection{Ground-Truth Statistics}  

From \textbf{17,047 table pairs}, after labeling our ground-truth consists of 52\% table pairs marked with \texttt{noalignment} relation, 24\% marked with as having \texttt{equivalent} alignment, and the remaining 23\% with \texttt{subPartOf} relation. The 47\% portion of table pairs with a relation,  result from \textbf{876 article pairs}, which presents a further $\blacktriangledown 4.2\times$ reduction of article pairs from our initial filtering step.

The \emph{average agreement rate} amongst crowdworkers for table pairs is 0.91, which is measured as a combination of the worker's confidence score and the agreement rate.

\subsection{Baselines and TableNet setup}\label{subsec:baselines}
We compare \emph{TableNet} in two main aspects: (i) efficiency in candidate pair generation, and (ii) table alignment.

\subsubsection{Candidate Generation Baselines} 
In the candidate generation phase we first look for article pairs whose tables are likely to yield an alignment $r(t_i, t_j)\neq\emptyset$. 

\noindent\textbf{Greedy -- G.} For each article we consider as pairs all other articles containing a table. It has maximal coverage, however the amount of irrelevant pairs is extremely high.

\noindent\textbf{Direct Categories -- C1.} We consider as pairs articles that are associated with the same \emph{directly connected} categories. Due to the noisy article-category associations, there is no guarantee that we will have maximal coverage of relevant pairs.

\noindent\textbf{Deepest Category -- C2.} Wikipedia articles are associated with categories that belong to different levels in the category hierarchy. As pairs we consider all articles that belong to the \emph{deepest category} in the hierarchy in $C$.

\noindent\textbf{Parent Categories -- PC.} To increase the coverage of relevant pairs, we consider as pairs, articles that have the same \emph{parent categories} based on their directly associated categories.

\noindent\textbf{Milne-Witten -- MW.} In MW we consider as pairs all articles that are related (for some threshold $\tau$) based on the Milne and Witten relatedness score~\cite{DBLP:conf/cikm/MilneW08}. We compute the relatedness score on the Wikipedia anchor graph.

\subsubsection{Table Alignment Baselines} We consider the following baselines for the table alignment step.

\textbf{Google Fusion.} The work in \cite{DBLP:conf/sigmod/SarmaFGHLWXY12} finds related tables for a given table by computing a set relatedness scores against all possible table candidates. Two tables are  related if their schemas are related based on max-weight bipartite graph matching score (see Section~\ref{sec:relatedwork} for a detailed discussion).
Google Fusion is unsupervised, thus, we use a threshold $\tau$ (we fine tune $\tau$ s.t we find the best F1 score) to classify table pairs as either having a relation or not.

\textbf{TableNet$_{LR}$}. Here, we consider as a competitor a standard supervised model based a logistic regression model, which we train using the features in Table~\ref{tbl:candidate_feature_list}. Here, our aim is to show the necessity of using more computationally heavy approaches like RNNs.

\textbf{LSTM and BiLSTM.} We use standard long-term short-memory networks to train a strong baseline for table alignment. Similarly, as in TableNet, here too, we will use the different column representations introduced in Section~\ref{subsec:table_alignment}. Similarly, we use a bidirectional LSTM as a baseline. 

\paragraph*{Setup: LSTM, BiLSTM \& TableNet.} We set the number of dimensions for the \emph{hidden layer} to be 100. We train the models for 50 epochs and use 60\% of the data for training, 10\% for validation, and the remaining 30\% for testing.

We represent columns based on three main representations, which we explained in Section~\ref{subsec:table_alignment}. The simplest representation is based on the column description which we mark with TableNet$^{desc}$, and then incrementally add the instance-value representation which we denote with TableNet$^{+val}$, and finally add the \emph{type} representation denoted with  TableNet$^{+type}$. In the cases where we represent a column through more than one representation, we simply add up the different representations. Similarly are represented the baselines LSTM and BiLSTM.

For classification we use the \emph{softmax} function, and optimize the models to minimize the \emph{categorical cross-entropy} loss.

\subsection{Evaluation Metrics}\label{subsec:eval_metrics}

We distinguish between two sets of evaluation metrics, aimed at measuring the performance of the candidate generation process, and the table alignment.

\textbf{Candidate Generation.} The main aim is to \emph{minimize} the amount of \emph{irrelevant article pairs} $\langle a_i, a_j\rangle = \emptyset$, and at the same time retain pairs whose tables have an alignment. We compute $\Delta$ as the metric measuring the amount of reduction we achieve w.r.t the \emph{greedy} approach in generating article pairs. 
\begin{equation}
	\Delta = 1 - \frac{\langle a_i, a_j\rangle}{k * |A|}\,\, \text{where } a_i \neq a_j \wedge a_i, a_j \in A
\end{equation}
where, $k$ is the number of source articles.

Coverage we measure through \emph{micro} and \emph{macro} recall indicated with $R_{\mu}$ and $R$, respectively. $R_\mu$ represents the recall in terms of all table pairs from all the source articles, whereas macro recall measures the average recall scores from all source articles. 

\textbf{Table Alignment.} We rely on standard evaluation metrics, such as \emph{precision} (P), \emph{recall} (R), and \emph{F1 score} (F1).

\section{Evaluation Results}\label{sec:evaluation}

In this section, we present in details the evaluation results for TableNet and our competitors in terms of candidate generation efficiency and coverage, and the performance in table alignment. 

\subsection{Candidate Generation}

Here we discuss the evaluation results in terms of efficiency in generating relevant article candidate pairs, and compare w.r.t $\Delta$  against \emph{greedy approach}. Additionally we show the recall scores in retaining relevant article pairs.

\paragraph*{Baselines} Table~\ref{tbl:baseline_candidate} shows the efficiency and coverage results for the baseline strategies. From the baselines we notice that the use of the Wikipedia category graph $\Psi$ reduces the amount of irrelevant pairs drastically. In terms of recall, baseline \textbf{PC} maintains high recall with $R=0.83$, and at the same time reduces by $\Delta=87\%$ the amount of irrelevant pairs when compared to \emph{greedy} approach.
 \textbf{MW} and \textbf{C2} achieve the highest reduction rate $\Delta$. However, for \textbf{MW} the coverage of relevant article pairs is very low with $R=0.49$. 

The results in Table~\ref{tbl:baseline_candidate} show that despite the high reduction rates for the different baselines, we still face the issue of either having a highly imbalanced ratio of relevant and irrelevant pairs, or in some cases like \textbf{C2} where the reduction rate is the highest, the recall is low $R=0.49$. Thus, the balance between coverage and efficiency is not maintained. We show that we can improve the deficiencies of the baseline approaches through our feature set in Table~\ref{tbl:candidate_feature_list}.

\begin{table}[ht!]
	\centering
	\scalebox{1.0}{
	\begin{tabular}{l l l l l}
		\toprule
		& $|\langle a_i, a_j\rangle|$ & $\Delta$ & \emph{rel. pairs} & $R$ \\ 
		\midrule
		
		\textbf{G} & 26,500,000 & - & 876 & 1.0\\
        \textbf{PC} & 3,486,031 & 0.87 & 724 & 0.83\\
		\textbf{C1} & 792,701 & 0.97 & 571 & 0.65\\
		\textbf{MW} & 33,890 & 0.99 & 429 & 0.49\\	
		\textbf{C2} & 6,738 & 0.99 & 440 & 0.50\\	
		\bottomrule
	\end{tabular}}
	\caption{Reduction rate for baselines. Higher $\Delta$ means that there are less irrelevant pairs for the table alignment step.}
	\label{tbl:baseline_candidate}	
\end{table}

\paragraph*{TableNet: Filtering \& Classification} We first filter out article pairs whose tables are unlikely to yield a relation and then classify the remaining pairs to further filter out irrelevant pairs.

\begin{figure}[ht!]
	\includegraphics[width=1.0\columnwidth]{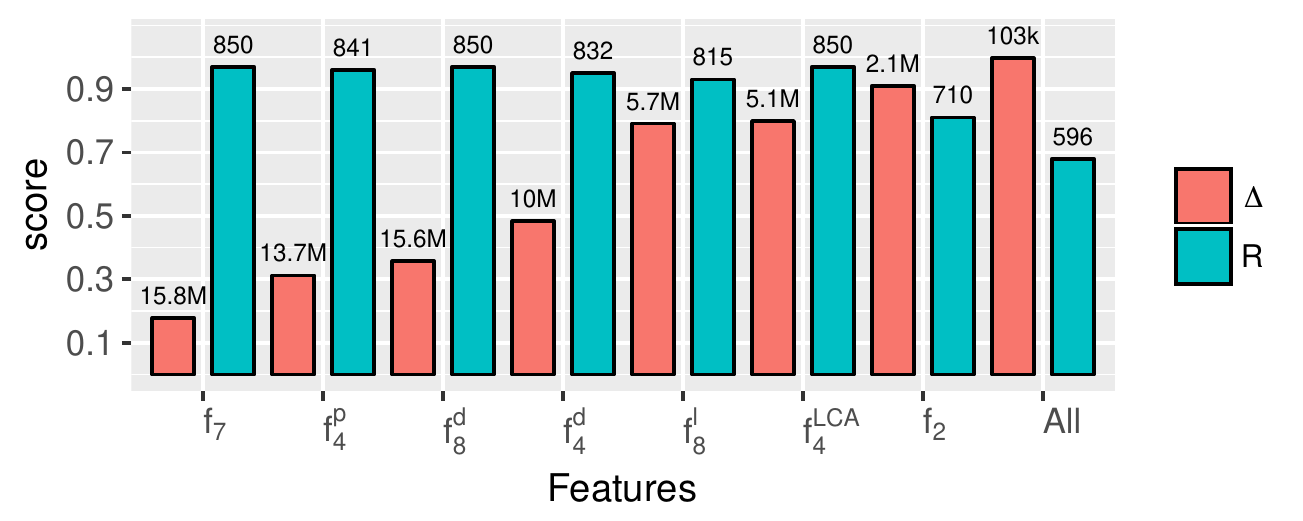}
	\caption{Feature impact in terms of reducing the amount of irrelevant pairs and the coverage of $\langle a_i, a_j\rangle_{r}$ pairs. On top of each bar we show the corresponding number of article pairs (for $\Delta$) and the number of retained relevant pairs (for $R$).}
	\label{fig:feature_filtering_impact}
\end{figure}

\textbf{Filtering. } The filtering step uses the features in Table~\ref{tbl:candidate_feature_list} to remove irrelevant article pairs. Figure~\ref{fig:feature_filtering_impact} shows the impact of the different features in reducing the amount of article pairs w.r.t \emph{greedy} approach. For instance, the $f_2$ feature, which computes the similarity of article abstracts based on their \emph{doc2vec} representation, provides a high reduction with $\Delta=0.91$. This feature follows our intuition on generating the article pairs for the ground-truth (see Section~\ref{sec:setup}), where the \emph{topic} and other semantic similarities for an article pair can be extracted from the article's content.

In terms of recall, we see that in majority of the cases the individual features have $R \geq 0.80$ coverage. 

Since different features capture different notions of similarity, we apply them in \emph{conjunction}, resulting in very high reduction rate of article pairs with $\Delta > 0.99$, and at the same time retaining a relatively high coverage with $R=0.68$. The reduction compared to the greedy approach is more than $\blacktriangledown 255$ times less pairs.

We believe that this high reduction rate and at the same time the relatively high recall of relevant pairs, when compared to the baseline approaches can be attributed to the fact that we consider the similarities of articles, and their corresponding categories and articles' content in the embedding space. This allows us to capture implicit semantics that cannot be capture for instance through the simple link-based structure in the category graph $\Psi$.

With the precomputed features we train a classifier and further filter out irrelevant pairs from the filtered articles in Figure~\ref{fig:feature_filtering_impact}.

\textbf{Classification.} Determining whether a pair of articles, specifically, if their corresponding tables will result in a alignment relation is a difficult learning task. 

From the previous filtering step, irrespective of the high reduction rate from 26M pairs to only 103k pairs, the amount of irrelevant pairs is still too high for any supervised approach to be able to learn models that predict with great accuracy the table relations. Thus, based on the configured RF model for high coverage (see Section~\ref{sec:setup}), we train it on the feature set in Table~\ref{tbl:candidate_feature_list} to further classify irrelevant pairs and filter them out.

\begin{figure}[h!]
	\centering
	\includegraphics[width=1.0\columnwidth]{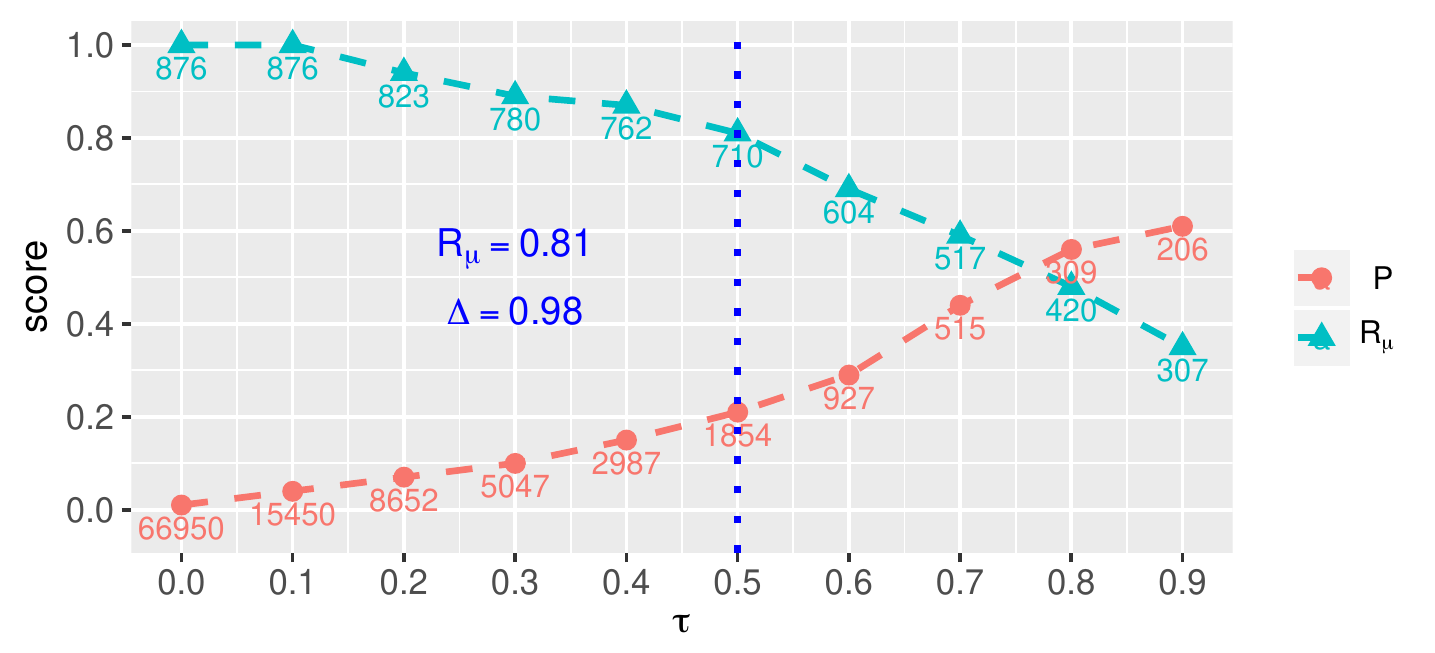}
	\caption{For varying classification confidence $\tau$ in the x-axis, we show the scores for $R_{\mu}$ (turquoise line), and the precision $P$ (red line). For each confidence score $\tau$ we show the corresponding amount of relevant pairs we retain in the case of recall, whereas for precision we show the amount of total pairs. We choose $\tau = 0.5$ which results in $R_\mu = 0.81$ and a $\Delta=0.982$ w.r.t the 103k pairs from the filtering step. }
	\label{fig:threshold_coverage}
\end{figure}

Figure~\ref{fig:threshold_coverage} shows the evaluation results for varying confidence thresholds of the RF model. With increasing threshold $\tau$ we can predict with higher accuracy pairs into their respective classes, whereas with lower thresholds we allow for more misclassifications. The increase of the confidence $\tau$ is directly proportional with the decrease in the amount of relevant pairs. This is intuitive as from the 103k pairs, only 876 pairs are actually relevant. However, based on the configuration of the RF (see Section~\ref{sec:setup}), we are able to retrieve relevant pairs with high coverage, by slightly allowing irrelevant pairs to pass through. 

We choose the confidence score to be $\tau = 0.5$, as it shows the best trade-off between the coverage of relevant pairs, and the amount of irrelevant pairs that are passed onto the table alignment step. We achieve a high reduction rate of $\Delta = 0.982$ leaving us with only 1.8k pairs, and with a recall of $R_\mu=0.81$.

Furthermore, if we compare this to the original amount of pairs from the greedy approach with 26M pairs, 103k from our pre-filtered candidates, and finally this drops to only 1.8k pairs in total after the classification step. Depending on the use case, one can use higher thresholds and thus have a higher ratio of relevant pairs, which makes the alignment task more efficient.

\subsection{Table Alignment}\label{subsec:table_alignment_evaluation}

\begin{table*}[ht!]
	\centering
	\scalebox{1.0}{
	\begin{tabular}{l l l l l l l l l l l l l}
	\toprule
	& \multicolumn{3}{c}{\texttt{equivalent}} & \multicolumn{3}{c}{\texttt{subPartOf}} & \multicolumn{3}{c}{\texttt{noalignment}} & & &\\
	\midrule
	& \textbf{P} & \textbf{R} & \textbf{F1} 	& \textbf{P} & \textbf{R} & \textbf{F1} & \textbf{P} & \textbf{R} & \textbf{F1} & \textbf{Acc} & \textbf{R} & \textbf{F1}\\
	\midrule
	\emph{Google Fusion} & 0.809	 & 0.405 & 0.540 & - & - &  - & - & - & -\\
	\emph{TableNet}$_{LR}$ & 0.824 & 0.790 & 0.804 &	 0.612 & 0.688 & 0.648 & 0.754 & 0.730 & 0.742 & 0.730 & 0.723 &  0.731 \\[1ex]

	\emph{LSTM}$^{desc}$  &	0.851 & \textbf{0.926} & 0.887 & \textbf{0.696} & 0.816 & 0.751 & 0.870 & \textbf{0.770} & 0.817 & 0.806 & 0.837 & 0.818\\
	\emph{LSTM}$^{+val}$ & 0.865 & 0.913 & 0.888 & 0.668 & 0.977 & 0.794 & 0.936 & 0.722 & 0.815 & 0.823 & 0.871 & 0.832\\		
	\emph{LSTM}$^{+type}$ &  0.839 & 0.935 & 0.884 & 0.547 & 0.976 & 0.701 & 0.933 & 0.564 & 0.703 & 0.773 & 0.825 & 0.763\\[1ex]

	\emph{BiLSTM}$^{desc}$ & 0.883 & 0.891 & 0.887 & 0.684 & 0.960 & 0.799 & 0.918 & 0.752 & 0.827 & 0.828 & 0.868 & 0.838\\
    \emph{BiLSTM}$^{+val}$ &0.877 & 0.871 & 0.874 & 0.684 & 0.975 & 0.804 & 0.915 & 0.747 & 0.823 & 0.826 & 0.864 & 0.834\\
\emph{BiLSTM}$^{+type}$ & 0.854 & 0.908 & 0.880 & 0.690 & 0.957 & 0.802 & 0.925 & 0.741 & 0.823 & 0.823 & 0.869 & 0.835\\[1ex]	

	\emph{TableNet}$^{desc}$ & \textbf{0.888} & 0.884 & 0.886 & 0.686 & 0.947 & 0.796 & 0.909 & 0.759 & 0.827 & 0.828 & 0.863 &  0.836\\	
	\emph{TableNet}$^{+val}$ & 0.856 & \textbf{0.926} & \textbf{0.890} &  0.675 & \textbf{0.993} & 0.804 &  \textbf{0.952} & 0.719 & 0.819 & 0.828 & \textbf{0.880} & 0.838\\
	\emph{TableNet}$^{+type}$ & 0.872 & 0.903 & 0.887 & 0.692 & 0.961 & \textbf{0.805} & 0.925 & 0.752 & \textbf{0.829} & \textbf{0.830} & 0.872 & \textbf{0.840} \\
	\bottomrule
	\end{tabular}}
	\caption{Evaluation results for the tasks of table alignment for the different competitors and TableNet. The evaluation results correspond to our manually constructed ground-truth dataset.}
	\label{tbl:alignment_results}
\end{table*}

In this section, we show the results for the table alignment step.  From the article pairs marked as relevant in the previous step, we classify the corresponding tables into their corresponding alignment relation, $r(t_i, t_j) \rightarrow$ \{\texttt{equivalent}, \texttt{subPartOf}, \texttt{none}\}.

\paragraph*{\textbf{Performance}} Table~\ref{tbl:alignment_results} shows the alignment evaluation results for TableNet and all the competitors. Apart from the \emph{Google Fusion} baseline, all the baselines are supervised models. In the case of \emph{Google Fusion}, we consider a table pair to be related if their matching score is above some threshold that we determine empirically s.t we have the highest F1 score. 

\textbf{Google Fusion.} This baseline has a reasonably high accuracy in determining whether a $r(t_i,t_j)\neq \emptyset$. Here we cannot distinguish between the different classes as the approach is unsupervised. In terms of recall it has the lowest score. This is due to the fact that the matching is performed by considering only the \emph{column type} and the \emph{column title}similarity. Additionally, the \emph{bipartite} matching algorithm cannot retain the order of the columns, which is highly important for determining the alignment relation.

\textbf{TableNet$_{LR}$.} In this baseline we trained a \emph{logistic regression} (LR) model with the feature set in Table~\ref{tbl:candidate_feature_list}, which classifies the \texttt{equivalent} and \texttt{subPartOf} relations with $F1=0.804$ and $F1=0.648$, respectively. When compared to Google Fusion, it achieves a 52\% relative improvement in terms of F1 score for \texttt{equivalent} class, whereas if we take the average of both classes \texttt{equivalent} and \texttt{subPartOf} then the F1 improvement is 37\%.

This shows that the proposed feature set is able to capture relations of type \texttt{equivalent} with high accuracy. However, it often misclassifies the \texttt{subPartOf} into the \texttt{none} and \texttt{equivalent} classes. One reason for this misclassification is since the features and the LR model cannot capture the sequence of columns, and it is not trivial to incorporate the information from the cell-values into the model for classifying the table relation.

\textbf{LSTM and BiLSTM.} One key motivation in this work is the hypothesis that through sequence based models, we can retain the order of columns in their respective schemas, an important aspect in determining the table alignment relation. The LSTM and BiLSTM approaches represent very competitive baselines. An additional advantage which addresses a deficiency in the standard supervised models, is that we jointly encode the different representations of a column for the classification task. Representing the columns as a combination of their description in the word embedding space, and the type and instance values through graph embeddings, we can capture complex relationship between the column description and the underlying cell-values.

For \texttt{equivalent} relations, LSTM$^{+val}$ and BiLSTM$^{+desc}$ achieve the highest F1 scores with $F1=0.886$ and $F1=0.887$, respectively. For \texttt{subPartOf} relations, the results look slightly different, with LSTM$^{+val}$ still having the highest F1 score, whereas for BiLSTM, BiLSTM$^{+val}$ the representation based on the column description and instance values achieves the highest F1 scores. The introduction of the column type in BiLSTM$^{+type}$ provides a further boost in the accuracy of determining \texttt{subPartOf} relations. One conclusion we draw from the comparison between the two relations and two models, is that for \texttt{subPartOf} relations the \emph{column type} provides additional power on determining the table alignment relation, whereas for \texttt{equivalent} it does not provide an additional advantage. These findings are inline with \cite{DBLP:conf/sigmod/SarmaFGHLWXY12}, where column type can provide important information in finding related tables. Comparing the LSTM and BiLSTM baselines against TableNet$_{LR}$, we gain 10\% relative improvement in terms of F1 score for \texttt{equivalent} relation, and with  22\% in the case of \texttt{subPartOf} relation. While for \emph{Google Fusion} we have a 64\% improvement for \texttt{equivalent} relation.

\textbf{TableNet.} In our approach, we addressed several deficiencies from the related work. Through our \emph{column-by-column} attention mechanism, we can compute soft alignments between columns in the respective table schemas and thus take into account the position of the matching columns in the corresponding schemas. Additionally, the column representations allow us to capture the similarity between columns and the schema context in which they appear, and additionally the representation context based on their description, type and its instance values. 

The evaluation results reflect this intuition. Comparing our best performing setup, TableNet$^{+type}$ achieves an overall $F1=0.840$ across all three classes. We achieve a relative improvement of 64\% when comparing F1 scores for the \texttt{equivalent} class against \emph{Google Fusion}, or 56\% if we compare the average F1 score for both alignment relations (\texttt{equivalent} and \texttt{subPartOf}). Against TableNet$_{LR}$ we observe high improvements for both alignment relation classes with a relative increase of 10\% and 24\% in terms of F1 score, for \texttt{equivalent} and \texttt{subPartOf}, respectively. 

LSTM and BiLSTM are two strong competitors. They are able to capture the sequence information in the table schemas, and additionally provide the means to capture the contextual similarity between the column description, type and instance cell-values. TableNet$^{+type}$ outperforms both approaches on average F1 score across all classes. For the individual classes, we note a variations amongst the different configurations of TableNet that perform best (marked in bold). The relative improvements are not as high as when compared against Google Fusion and TableNet$_{LR}$, however, they are consistent in nearly all cases. This confirms the usefulness of the attention mechanism for the alignment task, where we achieve an overall better performance in terms of F1 score.

\section{Conclusions and Future Work}\label{sec:conclusions}

In this work, we presented TableNet, an approach for table alignment by taking into account the coverage of table relations by providing an efficient approach for generating article pairs, whose tables we consider for alignment with a high accuracy. We provide fine-grained relations \texttt{equivalent}, \texttt{subPartOf}, or \texttt{none}, a significant improvement over existing works.

We constructed an exhaustive ground truth for a random sample of 50 Wikipedia articles for which we labeled all possible table pairs, providing a dataset against which we can measure the coverage of  table relations, and additionally provide high quality labels for more than 17k table pairs in our ground-truth. 

In  terms of \emph{efficiency}, we show that from a naive approach which produces 26.5M pairs we can provide an efficient means that guarantees a high coverage of more than 68\% and at the same time reducing the amount of pairs by $\blacktriangledown 255$ times. In terms of \emph{table alignment}, we show that we can improve over strong baselines. We showed relative improvement of 56\% when compared to \emph{Google Fusion}, and with 17\% when compared against TableNet$_{LR}$, a standard feature based model. If we compare against LSTM$^{+type}$ and BiLSTM$^{+type}$, we again achieve improvements in terms of F1 score, thus, validating our hypothesis that a \emph{column-by-column} attention mechanism provides soft alignments for columns across table schemas.

\paragraph*{Future Work} As future work we foresee the task of \emph{relation typing} s.t we can provide a attribute-based explanation of the relations in the case \texttt{equivalent} alignment, and for \texttt{subPartOf} provide attribute restrictions, i.e., in terms of the semantics of the respective tables, or in the form of \emph{selection} criteria that may apply to generate a sub or superset from a table s.t. the \texttt{subPartOf} alignment holds.

\end{document}